\documentclass[reprint,aps,nofootinbib,nobalancelastpage]{revtex4-2}
\usepackage{amsmath}
\usepackage{amsfonts}
\usepackage{amssymb}
\usepackage{amsthm}
\usepackage{mathtools}
\usepackage{graphicx}
\usepackage[dvipsnames]{xcolor}
\usepackage[colorlinks=True,citecolor=myBlue,linkcolor=myRed,urlcolor=myBlue,hypertexnames=false]{hyperref}
\usepackage{hypcap}
\usepackage{yfonts}
\usepackage{bm}
\usepackage[normalem]{ulem}
\usepackage{dsfont}
\usepackage{braket}
\usepackage{float}
\usepackage{marginnote}
\usepackage{tikz}
\usepackage{quantikz}
\usetikzlibrary{shapes, arrows.meta,decorations.pathreplacing}

\newcommand\cbox[1]{\vcenter{\hbox{#1}}}
\newcommand\rbar{\overline{R}}
\newcommand\delk{{\Delta^{(k)}}}
\newcommand\puk{{\cal P}^{(k)}}
\newcommand\lrb{L_{\overline{R}}}
\newcommand{\norm}[1]{\lVert#1\rVert}
\usepackage{kantlipsum}

\newcommand\eq[1]{\begin{align}#1\end{align}}

\newcommand\mytitle{Locality of deep thermalisation through the lens of entanglement teleportation}

\definecolor{myBlue}{RGB}{31,119,180}
\definecolor{myOrange}{RGB}{255,127,14}
\definecolor{myGreen}{RGB}{44,160,44}
\definecolor{myRed}{RGB}{214,39,40}
\definecolor{myPurple}{RGB}{148,103,189}

\begin{document}

\title{\mytitle}

\author{Saptarshi Mandal}
\email{saptarshi.mandal@icts.res.in}
\affiliation{International Centre for Theoretical Sciences, Tata Institute of Fundamental Research, Bengaluru 560089, India}

\author{Alan Sherry}
\email{alan.sherry@icts.res.in}
\affiliation{International Centre for Theoretical Sciences, Tata Institute of Fundamental Research, Bengaluru 560089, India}

\author{Sthitadhi Roy}
\email{sthitadhi.roy@icts.res.in}
\affiliation{International Centre for Theoretical Sciences, Tata Institute of Fundamental Research, Bengaluru 560089, India}

\begin{abstract}
Deep thermalisation characterises the emergence of universal quantum state ensembles on subsystems due to projective measurements on their complement. We study the notion of locality, or lack thereof, in this phenomenon by considering a subsystem partitioned into two disjoint subregions which remain causally disconnected at all times under unitary dynamics. We show that the onset of deep thermalisation in this geometry is fundamentally bounded by measurement-induced entanglement teleportation between the subregions. While measurements on the environment generate entanglement across the disconnected partitions -- suggesting an apparent non-locality -- we demonstrate that generic locally interacting systems exhibit an emergent locality. Specifically, the timescales for both deep thermalisation and entanglement teleportation scale logarithmically with the distance separating the subregions. Exceptions to this include special circuits where the randomness of the measurement outcomes is perfectly transmitted to the ensemble of states of the subsystem, conditioned on the outcomes; in
such cases the timescale for deep thermalisation is finite leading to genuine non-locality.
\end{abstract}

\maketitle

\paragraph*{Introduction:} The unitarity of quantum dynamics in isolated systems places fundamental bounds on where quantum information can spread and how fast.
In locally interacting systems, this is manifested in Lieb-Robinson bounds~\cite{lieb1972finite} and causal lightcones of information scrambling~\cite{bohrdt2017scrambling,luitz2017information,nahum2018operator,keyserlingk2018operator,khemani2018velocity} as well as in thermalisation timescales~\cite{deutsch1991quantum,srednicki1994chaos,rigol2008thermalisation,dalessio2016from,deutsch2018eigenstate}. 
An extreme example of this is presented by the setting,
\eq{
\cbox{\begin{tikzpicture}[scale=0.75]
\foreach \x in {-2,...,2}{
    \draw[thick] (\x,0) -- (\x,2.3);
    \draw[thick] (\x-0.15,0) -- (\x+0.15,0) -- (\x,-0.15) -- cycle;
}
\filldraw[fill=gray!50, draw=black, thick, rounded corners] (-2.2,0.2) rectangle (-0.8,0.7);
\filldraw[fill=gray!50, draw=black, thick, rounded corners] (0.8,0.2) rectangle (2.2,0.7);
\filldraw[fill=gray!50, draw=black, thick, rounded corners] (-1.2,1) rectangle (1.2,2);
\node at (0,1.5) {\small $U_t$};
\node at (-1.5,0.45) {\small $ W_1$};
\node at (1.5,0.45) {\small $ W_2$};
\node at (-2,-0.5) {\small $ R_1$};
\node at (-1,-0.5) {\small $ S_1$};
\node at (2,-0.5) {\small $ R_2$};
\node at (1,-0.5) {\small $ S_2$};
\node at (0,-0.5) {\small $ E$};
\end{tikzpicture}}\,,
\label{eq:setting}
}
where $W_1$, $W_2$ and $U_t$ are unitary operators, the initial states of subsystem $X$ will be denoted by $\ket{\phi_{X}}$ and $U_t$ can be thought of as the time-evolution operator.
In this setting irrespective of how efficiently scrambling the unitary $U_t$ is, the state of $R\equiv R_1\cup R_2$ is always given by
$\rho_R = \rho_{R_1}\otimes\rho_{R_2}$,
where
\eq{
\rho_{R_i} = {\rm Tr}_{S_i}[W_i\ket{\phi_{R_i}}\ket{\phi_{S_i}}\bra{\phi_{R_i}}\bra{\phi_{S_i}} W_i^\dagger]\,.
\label{eq:rho_Ri}
}
The upshot of the above is that 
no quantum correlations ever build up between $R_1$ and $R_2$ and $R$ is therefore never described via an ergodic, thermal ensemble.
In other words, the subsystem $\overline{R}\equiv S_1\cup E\cup S_2$ fails to act as a bath for $R$ simply because the architecture of the circuit makes $R_1$ and $R_2$ causally disconnected.

However, instead of tracing out the bath $\overline{R}$, performing projective measurements on it and generating an ensemble of conditional states on $R$ leads to a fundamentally new paradigm. 
Physically, this retains information about the bath encoded in the {\it projected ensemble}~\cite{cotler2023emergent,choi2023preparing},
\eq{
{\cal E}_{R}\equiv \{p(o_{\rbar}),\ket{\psi_R(o_{\rbar})}\}\,,
\label{eq:pe}
}
where $p(o_{\rbar})$ is the probability of obtaining the measurement outcome $o_{\rbar}$ and $\ket{\psi_R(o_{\rbar})}$ is the corresponding state.
The $k^{\rm th}$ moment of the ensemble is defined as
\eq{
\rho_R^{(k)} = \sum_{o_{\rbar}}p(o_{\rbar})[\ket{\psi_R(o_{\rbar})}\bra{\psi_R(o_{\rbar})}]^{\otimes k}\,,
\label{eq:pe-moments}
}
where the $\rho_R^{(1)}=\rho_R$ is simply the reduced density matrix of $R$.
Moments with $k\geq 2$ contain more information which cannot be reconstructed from $\rho_R$.
This has led to the concept of {\it deep thermalisation}~\cite{cotler2023emergent,choi2023preparing,ippoliti2022solvablemodelofdeep,ho2022exact,ippolitu2023dynamical,mark2024maximum,chang2025charge,varikuti2024unravelingemergence,manna2025peconserved,yu2025mixed,sherry2026deep} wherein the entire ensemble approaches a universal maximum-entropy ensemble subject to the constraints in the dynamics. In the absence of any conservation laws this is simply the Haar ensemble which leads to ${\cal E}$ forming a design. 

A key point to note in this setting is that the measurements on ${\rbar}$ manifestly break the unitarity of the dynamics, which in turn renders moot the constraints placed by unitary dynamics.
This raises the fundamental question that is there an emergent non-locality in deep thermalisation effected by the non-unitarity of the measurement-unitary hybrid dynamics?
In this work we address this question in a setting illustrated in \eqref{eq:setting} through the lens of entanglement teleportation between $R_1$ and $R_2$.
The conceptual premise is that for the ensemble ${\cal E}$ to thermalise deeply, the states in ${\cal E}$ must carry quantum correlations between $R_1$ and $R_2$ betrayed by entanglement between them. In fact, we show concretely that the timescales for deep thermalisation are bounded by the entanglement teleportation timescales. 

Using this idea we show that there is perfect non-locality of deep thermalisation and entanglement teleportation if the inherent randomness of the measurement outcomes is transmitted perfectly through the unitary channels formed by $U_t$, $W_1$ and $W_2$. 
This is the case if these unitaries are maximally entangling such as circuits with Haar-random gates with infinitely large local Hilbert-space dimension or with dual-unitary gates.
On the other hand, we find that in more generic settings, the non-locality is suppressed and there exists a timescale for deep thermalisation which grows logarithmically in the separation between $R_1$ and $R_2$.
This constitutes the main result of this work.

In the remainder of the paper, we first prove an inequality, quite generally, between timescales of deep thermalisation of a subsystem and those of entanglement teleportation between its constituents.
As a proof of principle, we then show for a toy model of $U_t$ how measurements on $\overline{R}$ lead to deep thermalisation of $R$ and entanglement teleportation between $R_1$ and $R_2$ even though they are not causally connected. 
We then present results for $U_t$ described by a locally interacting circuit where teleportation leads to an effective locality in deep thermalisation manifested in timescales that grow with the separation between $R_1$ and $R_2$. 
We provide an understanding of these results via a heuristic picture based on random tensor networks (RTNs).

For simplicity, we will consider, $W_i\ket{\phi_{R_i}\phi_{S_i}}$ to be the even-parity Bell state,
which implies $\rho_{R_i}\propto \mathbb{I}$.
For this choice, if the ensemble ${\cal E}$ thermalises deeply, it would do so to the Haar ensemble. 
For other choices, the deep thermal ensemble would be the Scrooge ensemble~\cite{mark2024maximum,chang2025charge,mcginley2025scroogeensemblemanybodyquantum,mok2026naturestingyuniversalityscrooge} generated by $\rho_R$ in Eq.~\eqref{eq:rho_Ri}.

\paragraph*{Bound on deep thermalisation timescales:}We first show how the dynamics of deep thermalisation is bounded by entanglement teleportation.
As a measure of the distance between the $k^{\rm th}$ moments of ${\cal E}$ and the Haar ensemble we consider
\eq{
\delk \equiv \norm{\rho_R^{(k)}-\rho_{\rm Haar}^{(k)}}_2\,,
\label{eq:Deltak}
}
where $\norm{\cdot}_2$ denotes the Hilbert-Schmidt norm~\footnote{We choose this norm purely for analytical convenience; the physics remains unchanged for other choices of the norm.}.
To quantify the teleported entanglement in ${\cal E}$, we define the average $k^{\rm th}$ bipartite purity between $R_1$ and $R_2$ as
\eq{
\puk = \sum_{o_{\rbar}}p(o_{\rbar}){\rm Tr}_{R_1}[{\rm Tr}_{R_2} \ket{\psi_R(o_{\rbar})}\bra{\psi_R(o_{\rbar})}]^k\,.
\label{eq:purity-k}
}
Specifically, it will be relevant to compare the averaged purity relative to the Haar ensemble, $\delta\puk = \puk-\puk_{\rm Haar}$, as this directly encodes how `Haar-like' are the quantum correlations between $R_1$ and $R_2$ in ${\cal E}$.
The purity difference can be expressed as 
\eq{
\delta \puk = {\rm Tr}_{R^{\otimes k}}\left[(\mathbb{S}_{R_1^{\otimes k}}
\otimes
\mathbb{I}_{R_2^{\otimes k}}
)
(\rho_{R}^{(k)}-\rho^{(k)}_{\rm Haar})\right]\,,
}
where $\mathbb{S}_{R_1^{\otimes k}}$ denotes the cyclic permutation operator acting on the $k$ replicas of $R_1$ and ${\rm Tr}_{R^{\otimes k}}$ denotes the trace over the $k$ replicas of $R$.
Using the Cauchy-Schwarz inequality for operator inner products we have
\eq{
\delta \puk \leq \norm{\mathbb{S}_{R_1^{\otimes k}}
\otimes
\mathbb{I}_{R_2^{\otimes k}}}_2
\norm{\rho_{R}^{(k)}-\rho^{(k)}_{\rm Haar}}_2\,,
}
which using the definition in Eq.~\eqref{eq:Deltak}, leads to 
\eq{
\delk\geq D_R^{-k/2} \delta \puk\,,
\label{eq:delk-puk-ineq}
}
where $D_R\!=\!D_{R_1}D_{R_2}$ is the Hilbert-space dimension of $R$.
The physical import of the inequality is that for the $k^{\rm th}$ moment of ${\cal E}$ to be arbitrarily close to that of the Haar ensemble, the purity difference $\delta \puk$ must also be arbitrarily small.
In other words, sufficient entanglement must be teleported between $R_1$ and $R_2$ such that it is arbitrarily close to the Page value.
Therefore, timescales associated with entanglement teleportation manifestly place lower bounds on those for deep thermalisation. 

\paragraph*{Minimal model:} As a proof of principle to demonstrate that subsystems with disjoint, causally disconnected constituents can thermalise deeply accompanied by entanglement teleportation, we discuss a minimal, toy model devoid of any spatial structure~\cite{ippoliti2022solvablemodelofdeep}. 
In particular, we show that in this model, deep thermalisation of $R$ and entanglement teleportation between $R_1$ and $R_2$ can happen within $O(1)$ timescales.

The model is of the form shown in Eq.~\eqref{eq:setting} where the bath $E$ is tripartitioned into $E_1$, $E_2$ and $E_3$ such that $S_{1/2}$ interacts only with $E_{1/2}$.
Pictorially, $U_t$ can be represented as 
\eq{
    \cbox{
\begin{tikzpicture}[scale=0.75]
\foreach \x in {-1,...,1}{
    \draw[thick] (\x,0.5) -- (\x,2.5);
}
\filldraw[fill=gray!50, draw=black, thick, rounded corners] (-1.2,1) rectangle (1.2,2);
\node at (0,1.5) {\small $U_{t=2}$};
\node at (-1,0.0) {\small $S_1$};
\node at (1,0.0) {\small $S_2$};
\node at (0,0.0) {\small $E$};
\end{tikzpicture}
}
=
\cbox{
\begin{tikzpicture}[scale=0.75]
\foreach \x in {-2,...,2}{
    \draw[thick] (\x,0.5) -- (\x,3.0);
}
\foreach \y in {1,2}{
\filldraw [fill=YellowOrange, draw=black, thick,rounded corners] (-1.2,\y-0.2) rectangle (1.2,\y+0.2);}
\foreach \y in {1,2}{
\filldraw [fill=NavyBlue!50, draw=black,thick,rounded corners] (-2.2,\y-0.2+0.5) rectangle (-0.8,\y+0.2+0.5);
\filldraw [fill=NavyBlue!50, draw=black,thick,rounded corners] (0.8,\y-0.2+0.5) rectangle (2.2,\y+0.2+0.5);}
\draw[Blue,thick,dashed,rounded corners]
(-2.5,1.75)--++(0,-1.07)--++(5.0,0)--++(0,1.07)--cycle;
\node at (-2,0.25) {\small $S_1$};
\node at (-1,0.25) {\small $E_1$};
\node at (0,0.25) {\small $E_3$};
\node at (1,0.25) {\small $E_2$};
\node at (2,0.25) {\small $S_2$};
\node at (0,1) {\small $u_{E,1}$};
\node at (0,2) {\small $u_{E,2}$};
\end{tikzpicture}
}\,,
    \label{eq:toy-model-unitary}
}
where the dashed box denotes the time-evolution operator over one time step.
We will consider the Hilbert-space dimension $D_{E_3}\to \infty$ but restrict $D_{E_{1/2}}=2$.
In addition, we consider the gates $\{u_{E,t}\}$ in Eq.~\eqref{eq:toy-model-unitary} acting on $E$ to be independent Haar-random unitaries.
This construction ensures that $E$ is a `perfect' bath, but interacts with the rest of the system through bottlenecks, characteristic of locally interacting systems. 

More importantly, this construction also allows one to replace the ensemble over the measurement outcomes in $E$ by a Haar-random state in the space of bonds that connect $S_{1/2}$ to $E_{1/2}$~\cite{ippoliti2022solvablemodelofdeep}.
Diagrammatically, an unnormalised state in the ensemble can be represented as
\eq{
    \ket{\tilde{\psi}_R(z_S,\phi)} = \cbox{\begin{tikzpicture}[scale=0.75]
\foreach \x in {-4,-3,1.1,2.1}{
\draw[thick] (\x,0)--(\x,3.5);
}
\foreach \x in {-4,-3,1.1,2.1}{
\filldraw[fill=white,draw=black,thick] (\x-0.15,0)--(\x+0.15,0)--(\x,-0.2)--cycle;
}
\foreach \x in {-3,1.1}{
\filldraw[fill=Green!40,draw=black,thick] (\x-0.15,3.5)--(\x+0.15,3.5)--(\x,3.7)--cycle;
}
\foreach \i in {0,1}{
\draw[thick,rounded corners] (-2,1.5+\i)--++(0,0.3)--++(0.3+\i*0.6,0)--++(0,-1.5-\i-0.3);
\draw[thick,rounded corners] (-2,1.5+\i)--++(0,-0.3)--++(\i*0.6,0)--++(0,-1.2-\i);
\draw[thick,rounded corners] (0.1,1.5+\i)--++(0,0.3)--++(-0.3-\i*0.6,0)--++(0,-1.5-\i-0.3);
\draw[thick,rounded corners] (0.1,1.5+\i)--++(0,-0.3)--++(-\i*0.6,0)--++(0,-1.2-\i);
}
\filldraw[draw=black,fill=Red!10,thick] (-2.2,0)--(0.3,0)--(-0.9,-0.6)--cycle;
\foreach \y in {1,2}{
\filldraw[fill=NavyBlue!50,draw=black,rounded corners,thick] (-3.2,\y-0.2+0.5) rectangle (-1.8,\y+0.2+0.5);
\filldraw[fill=NavyBlue!50,draw=black,rounded corners,thick] (-0.1,\y-0.2+0.5) rectangle (1.3,\y+0.2+0.5);
}
\filldraw[fill=gray!50,draw=black,rounded corners,thick] (-4.2,0.3) rectangle (-2.8,0.7);
\filldraw[fill=gray!50,draw=black,rounded corners,thick] (0.9,0.3) rectangle (2.3,0.7);
\draw[red,thick,dashed,rounded corners] (-0.9,3.7)--++(-3.4,0)--++(0,-3.75-0.3)--++(1.75,0)--++(0,1)--++(4.3-2.7,0);
\draw[red,thick,dashed,rounded corners]
(-0.7,3.7)
--++(3.1,0)
--++(0,-3.75-0.3)
--++(-1.75,0)
--++(0,1)
--++(-1.6,0);
\draw[thick] (-4,3.4)--++(0,0.5);
\draw[thick] (2.1,3.4)--++(0,0.5);
\node at (-0.9,-0.25) {\small $\ket{\phi}$};
\node at (-2.9,4.0) {\small $\ket{z_{S_1}}$};
\node at (1.2,4.0) {\small $\ket{z_{S_2}}$};
\end{tikzpicture}}\,,
    \label{eq:psi-haar-state}
}
where $z_S = z_{S_1}\otimes z_{S_2}$ labels the measurement outcome on $S_1\cup S_2$ and $\ket{\phi}$ is a Haar-random state of dimension $D_\phi = (D_{E_1}D_{E_2})^{2t}$.
Using the above, the $k^{\rm th}$ moment of the ensemble, defined in Eq.~\eqref{eq:pe-moments}, can be computed as
\eq{
\rho_{R}^{(k)}
=
\sqrt{D_\phi}
\int_{\rm Haar} \!\!\!\!d\phi~
\sum_{z_S}
\frac{
\left(
{\cal L}_{z_S}\ket{\phi}\bra{\phi}{\cal L}^{\dagger}_{z_S}
\right)^{\otimes k}
}{
\braket{\phi|{\cal L}^{\dagger}_{z_S}{\cal L}_{z_S}|\phi}^{k-1}
}\,,
\label{eq:toy-model-moment}
}
where ${\cal L}_{z_S}$ is a linear operator that acts on $\ket{\phi}$ and outputs a state on $R$; it is denoted by the red dashed box in Eq.~\eqref{eq:psi-haar-state}.
The result above suggests that at any time $t$, ${\cal E}_R$ is described by a maximum-entropy, generalised Scrooge ensemble~\cite{mark2024maximum} 
which is a mixture of several Scrooge ensembles each with its own probability.
Specifically, 
% \eq{
    $\rho_R^{(k)} = \sum_{z_S}p(z_S)\rho_{\rm Scrooge}^{(k)}[\varrho(z_S)]\,,$
% }
where $\varrho({z_S})$ and $p(z_S)$ are given by
\eq{
\varrho(z_S)= \frac{{\cal L}_{z_S}{\cal L}^\dagger_{z_S}}{{\rm Tr}[{\cal L}_{z_S}{\cal L}^\dagger_{z_S}]}\,,~~p(z_S) = \frac{{\rm Tr}[{\cal L}_{z_S}{\cal L}^\dagger_{z_S}]}{\sqrt{D_\phi}}\,.
}
The emergence of a maximum-entropy ensemble for ${\cal E}_R$ at any time $t$ is a manifestation of the fact that the model, by construction, has a perfect bath and hence the randomness of the measurements is transmitted optimally to the bottleneck. 
In general, this randomness is not transported further optimally.
However, even if one of the gates between $S_1$ and $E_1$, and similarly between $S_2$ and $E_2$ can transmit the randomness perfectly, ${\cal E}_R$ immediately forms the Haar ensemble and $\puk$ takes the corresponding Page value.
The simplest such example of a perfectly transmitting gate is a dual-unitary gate; the details of this scenario are presented in the Supplementary Material (SM)~\cite{supp}. 

\begin{figure}
\includegraphics[width=\linewidth]{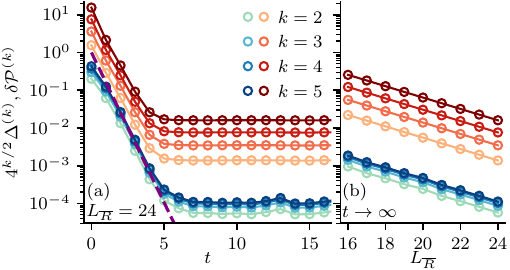}
\caption{Numerical results for the minimal model in Eq.~\ref{eq:toy-model-unitary}. (a) $\delk$ (rescaled by $4^{k/2}$) and $\delta\puk$ as a function of $t$ for different values of $k$ with the former in red tones and the latter in blue tones. The data show an exponential decay with $t$ where both the quantities have the same rate which is independent of $k$. The purple dashed line shows $\sim e^{-\gamma t}$ with $\gamma = \ln(25/4)$. The Hilbert-space dimension $D_{E_3}$ is parametrised as $D_{E_3}=2^{L_{\overline{R}}-4}$. The data are for $L_{\overline{R}}=24$ and $D_{E_{1/2}}=2=D_{S_{1/2}}$. (b) The saturation of the data is a finite $D_{E_3}$ effect as the saturation value for both quantities decays exponentially with $L_{\overline{R}}$.  }
\label{fig:toy-model-plot}
\end{figure}

As a representative of a more general case, we consider all the gates between $S_{1/2}$ and $E_{1/2}$ to be Haar-random gates. 
Results for this case, averaged over the Haar-randomness of the gates, are shown in Fig.~\ref{fig:toy-model-plot} which are consistent with the inequality in Eq.~\eqref{eq:delk-puk-ineq}.
The data show that both $\delk$ and $\delta\puk$ decay exponentially with $t$: $\delk,\delta\puk\sim e^{-\gamma t}$ which provides evidence for the fact that in this minimal model, both deep thermalisation of $R$ with disjoint constituents $R_1$ and $R_2$, as well as entanglement teleportation between $R_1$ and $R_2$ happen within $O(1)$ timescales.
The data also shows that the rate $\gamma$ is the same for both  $\delk$ and $\puk$, independent of $k$ and to an excellent approximation, given by $\gamma \approx \ln(25/4)$.

The above features can be understood via the following heuristic argument. 
Averaging over the Haar-random gates between $S_{1/2}$ and $E_{1/2}$, as well as the Haar-random state $\ket{\phi}$, leads to an effective transfer matrix, repeated applications of which eventually give $\delk$ and $\delta\puk$ at time $t$. 
The steady state of this process corresponds to the Haar ensemble, and the exponential approach to it is governed by the subleading eigenvalue of this transfer matrix.
This eigenvalue can be estimated using an argument based on operator scrambling. Consider one of the Haar-random gates acting on $S_1 \cup E_1$. The operator which remains invariant under the gate and contributes to the steady state is naturally $\mathbb{I}_{S_1}\otimes\mathbb{I}_{E_1}$. Any other product of Pauli operators is scrambled uniformly across the 15 remaining non-trivial Pauli strings on  $S_1\cup E_1$.
However, because the $E_1$ bonds are contracted with the Haar-random state $\ket{\phi}$ which is also averaged over, the only scattered strings that survive the average are those of the form $\Sigma_{S_1}\otimes\mathbb{I}_{E_1}$ where $\Sigma_{S_1}$ could be any of the three non-trivial Pauli operators.
There are exactly 3 such strings. The total relative weight that survives the contraction is therefore $\lambda = 2\times 3/15 = 2/5$, where the factor of 2 accounts for the trace over $\mathbb{I}_{E_1}$.
Since the gates between $S_2$ and $E_2$ are independent of those between $S_1$ and $E_1$, they independently contribute an identical factor of $\lambda$. This factorisation dictates that the subleading eigenvalue for the global transfer matrix is $\lambda^2 = 4/25$, which is precisely the rate $e^{-\gamma}$ seen in Fig.~\ref{fig:toy-model-plot}.

\begin{figure}
\includegraphics[width=\linewidth]{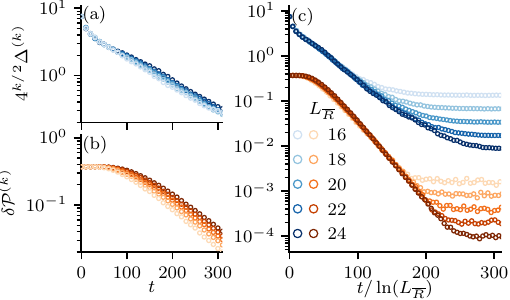}
\caption{Numerical results for (a) $\delk$ and (b) $\delta \puk$ with $k=4$ for the brickwork circuit described in Eqs.~\eqref{eq:brickwork}-\eqref{eq:ham}. Darker colours denote larger system sizes $\lrb$.
The data show that at $O(1)$ intermediate times $\delk$ and $\delta \puk$ grow with $\lrb$, indicating that the deep thermalisation timescale for the ensemble ${\cal E}_R$, which is constrained by entanglement teleportation between $R_1$ and $R_2$, grows with $\lrb$. (c) Scaling time, $t$, by $\ln\lrb$ collapses the data for different $\lrb$ onto a common curve indicating that the timescale scales as $t_\ast(\lrb)\sim\ln\lrb$.
}
\label{fig:ising-plot}
\end{figure}

\paragraph*{Locally interacting circuits:}
To address the emergence of locality in deep thermalisation, we consider a spatially extended system with local interactions. 
The locality of deep thermalisation in this case manifests itself in the fact that entanglement teleportation and deep thermalisation occur over a timescale that grows logarithmically with the separation between $R_1$ and $R_2$. 
We present numerical results to this end, and a justification of the logarithmic scaling using random tensor networks.
Specifically, we consider a random circuit in one spatial dimension with brickwork geometry,
\eq{
\cbox{
\begin{tikzpicture}[scale=0.75]
\foreach \x in {-1,...,1}{
    \draw[thick] (\x,0.5) -- (\x,2.5);
}
\filldraw[fill=gray!50, draw=black, thick, rounded corners] (-1.2,1) rectangle (1.2,2);
\node at (0,1.5) {\small $U_{t=2}$};
\node at (-1,0.1) {\small $S_1$};
\node at (1,0.1) {\small $S_2$};
\node at (0,0.1) {\small $E$};
\end{tikzpicture}
}
=
\cbox{
\begin{tikzpicture}[scale=0.75]
% vertical wires
\foreach \i in {1,...,8}{
    \draw[thick] (\i*0.75,0) -- (\i*0.75,2.5);
}
% right-moving brickwork (odd columns)
\foreach \i in {1,3,5,7}{
    \foreach \t in {1,3}{
        \filldraw[
        fill=NavyBlue!50,
        draw=black,
        rounded corners,
        thick]
        (\i*0.75-0.1*0.75,\t*0.5-0.2)
        rectangle
        (\i*0.75+1.1*0.75,\t*0.5+0.2);
    }
}
% left-moving brickwork (even columns)
\foreach \i in {2,4,6}{
    \foreach \t in {2,4}{
        \filldraw[
        fill=NavyBlue!50,
        draw=black,
        rounded corners,
        thick]
        (\i*0.75-0.1*0.75,\t*0.5-0.2)
        rectangle
        (\i*0.75+1.1*0.75,\t*0.5+0.2);
    }
}
% labels (left side renamed)
\node at (0.75,-0.25) {\small $S_1$};
\node at (1.5,-0.25) {\small $E_1$};
% right side renamed
\node at (7*0.75,-0.25) {$E_2$};
\node at (8*0.75,-0.25) {$S_2$};
\draw[decorate,decoration={brace,mirror,amplitude=5pt}]
(3*0.75,-0.2) -- ++(3*0.75,0)
node[midway,below=4pt] {\small $E_3$};
\draw[Blue,thick,dashed,rounded corners]
(0.5,1.25)--++(0,-1.07)--++(5.75,0)--++(0,1.07)--cycle;
\end{tikzpicture}
},
\label{eq:brickwork}
}
where each leg corresponds to a qubit and we will denote the distance between $R_1$ and $R_2$ by $\lrb$.
The gates of the circuit are chosen to be
\eq{
\cbox{
\begin{tikzpicture}[scale=0.75]
\foreach \x in {0,0.75}{
    \draw[thick] (\x,-0.35) -- (\x,0.35);
}
\filldraw[
    fill=NavyBlue!50,
    draw=black,
    rounded corners=3.5pt,
    thick
] (-0.075,-0.17) rectangle (0.75+0.075,0.17);
\end{tikzpicture}}
= [u_j\otimes u_{j+1}]\exp(-i \tau H)[v_j\otimes v_{j+1}]\,,\label{eq:gate}
}
where $\{u_j,v_j\}$ are single-qubit Haar-random gates and 
\eq{
\begin{split}
H = 0.3X_jX_{j+1} &+ 0.2(X_j + X_{j+1}) + \\&0.4Z_jZ_{j+1} + 0.5(Z_j + Z_{j+1})\,,
\end{split}
\label{eq:ham}
}
with $X_j (Z_j)$ denoting the Pauli-$X (Z)$ operator on site $j$ and we choose $\tau=0.28$.
As before, in the initial state we consider $R_{1/2}$ and $S_{1/2}$ to be in Bell-pair states and $E$ to be in a random product state over each of its sites. 
The numerical results, averaged over the initial states in $E$, as well as the Haar-random gates $\{u_j,v_j\}$ in Eq.~\eqref{eq:ham} are presented in Fig.~\ref{fig:ising-plot}. The results are consistent with the inequality in Eq.~\eqref{eq:delk-puk-ineq}.

In Fig.~\ref{fig:ising-plot}(a), we show the results for $\delk$ for $k=4$ (results for other values of $k$ are qualitatively similar and for the dependence on $k$, see the SM~\cite{supp}). 
At the earliest times, $\delk$ decays with $t$, independent of $\lrb$.
However, this does not signify the onset of deep thermalisation at any $O(1)$ time. 
This is made evident in Fig.~\ref{fig:ising-plot}(b) where the ensemble averaged purity between $R_1$ and $R_2$ remains arbitrarily close to its initial value of 1.
This implies that no quantum correlations or entanglement develops between $R_1$ and $R_2$ in this temporal regime.

More importantly, at any intermediate, finite time, $\delk$ and $\delta \puk$ grow with $\lrb$ indicating that the ensemble ${\cal E}_R$ is further away from the Haar ensemble for larger $\lrb$.
Equivalently, the timescale for deep thermalisation and generation of entanglement between $R_1$ and $R_2$ grows with $\lrb$. 
The results in Fig.~\ref{fig:ising-plot}(c) show that scaling $t$ by $\ln\lrb$ collapses the data of $\delk$ and $\delta \puk$ for different $\lrb$ onto a common `universal' curve.
This identifies a timescale, $t_\ast(\lrb)\sim \ln \lrb$, beyond which  $\delta \puk$ decays and entanglement is generated between $R_1$ and $R_2$ accompanied by the onset of deep thermalisation.

The emergence of this timescale manifestly implies the locality of deep thermalisation. 
The fact that $\delta \puk$ does not start decaying until $t_\ast(\lrb)\sim \ln\lrb$ together with the inequality in Eq.~\eqref{eq:delk-puk-ineq} naturally means that the onset of deep thermalisation must itself be constrained by a timescale which scales at least logarithmically with $\lrb$. 
We note that this scaling is consistent with bounds for entanglement teleportation in finite-depth circuits~\cite{bao2024finite}.
It is important to re-emphasise at this point that the entanglement between $R_1$ and $R_2$ in this case is purely due to the measurements on $S\cup E$ as the model is still of the form in Eq.~\eqref{eq:setting} and hence, purely unitary dynamics cannot generate any entanglement between $R_1$ and $R_2$.

In the following, we will show that the logarithmic in $\lrb$ scaling of $t_\ast$ can be complementarily and independently understood via quite general arguments based on RTNs~\cite{garnerone2010rmps,garnerone2011stat,garnerone2013rmps,haferkamp2021stat,lami2025designs,lami2025statedesign,lami2026rmpsdesigns}. 
This approach also has the practical advantage of accessing system sizes much larger than those whose brickwork circuit dynamics are feasible to simulate while also leading to analytical tractability.
The argument is based on the premise that states generated by scrambling local circuits, such as the model discussed above, can be proxied effectively by Gaussian RTNs where the depth of the circuit is encoded in the bond dimensions of the tensors.
Such an RTN state can be written as 
\eq{
    \ket{\psi_{\rm RTN}}\propto\cbox{
\begin{tikzpicture}[scale=1]
\draw[thick] (0,0) -- (2,0);
\draw[thick] (3.5,0) -- (4.5,0);
\draw[thick,dashed] (2,0) -- (4,0);
\foreach \x in {0,1,2,3.5,4.5}{
    \draw[black,thick] (\x,0) -- (\x,0.5);
    \filldraw[draw=black,thick,fill=Green!20, rounded corners] (\x-0.25,-0.25) rectangle (\x+0.25,0.25);
}
\node at (0,-0.5) {\small $M^{[0]}$};
\node at (1,-0.5) {\small $M^{[1]}$};
\node at (2,-0.5) {\small $M^{[2]}$};
\node at (3.5,-0.5) {\small $M^{[L]}$};
\node at (4.5,-0.5) {\small $M^{[L+1]}$};
\end{tikzpicture}
}\,,
    \label{eq:psi-rtn}
}
where the sites $0$ and $L+1$ can be identified with $R_1$ and $R_2$ respectively, sites $1$ through $L$ can be identified with $\overline{R}$.
Each rectangle $M^{[i]}$ for $i\in [1,L]$ is a $\chi_i\times\chi_{i+1}\times 2$ Gaussian random tensor where $\chi_{i(i+1)}$ is the bond dimension of its left(right) horizontal bond and the physical bond, denoted by the vertical legs has a bond dimension of 2 as the state is defined on qubits. 
The boundary tensors $M^{[0]}$ and $M^{[L+1]}$ have dimensions of $\chi_1\times 2$ and $\chi_{L+1}\times 2$ respectively. 

For a scrambling circuit which exhibits ballistic entanglement growth, the entanglement across the cut between sites $i$ and $i+1$ (for $i\leq L/2$) at depth $t$ is given by ${\cal S}_i(t)=\min[v_{\rm E}t, \ln D_i]$ where $v_{\rm E}$ is the entanglement velocity and $D_i=2^{i+1}$ is the Hilbert-space dimension of the subsystem to the left of the cut. 
Since $\{M^{[i]}\}$ are Gaussian random tensors, the Schmidt rank across the cut between sites $i$ and $i+1$ is well approximated by $\chi_i$.
Motivated by this, we assign ${\chi_{i}\!=\!\exp [{\cal S}_i(t)]}$.
For $i>L/2$, the profile of the bond dimensions is a mirror image of those for $i\leq L/2$.
As marginals of Gaussian random tensors are also Gaussian random tensors, projective measurements on sites $1$ through $L$ lead to a conditional state on sites $0\cup (L+1)$,
\eq{
\ket{{\tilde{\psi}}_{\rm RTN}}\!\propto\!\!\!\cbox{
\begin{tikzpicture}[scale=1]
\draw[thick] (0,0) -- (2,0);
\draw[thick] (3.5,0) -- (4.5,0);
\draw[thick,dashed] (2,0) -- (4,0);
\draw[black,thick] (0,0) -- (0,0.5);
\draw[black,thick] (4.5,0) -- (4.5,0.5);
\foreach \x in {0,1,2,3.5,4.5}{
    \filldraw[draw=black,thick,fill=Blue!10, rounded corners] (\x-0.25,-0.25) rectangle (\x+0.25,0.25);
}
\node at (0,-0.5) {\small $\tilde{M}^{[0]}$};
\node at (1,-0.5) {\small $\tilde{M}^{[1]}$};
\node at (2,-0.5) {\small $\tilde{M}^{[2]}$};
\node at (3.5,-0.5) {\small $\tilde{M}^{[L]}$};
\node at (4.5,-0.5) {\small $\tilde{M}^{[L+1]}$};
\end{tikzpicture}
},
\label{eq:rtn-post-meas}
}
where $\tilde{M}^{[i]}$ is a $\chi_i\times\chi_{i+1}$-dimensional Gaussian random tensor.
We numerically generate several instances of such RTN states and compute the bipartite purity between $R_1$ and $R_2$ in each of them as well as $\rho_R^{(2)}$ over the ensemble of the RTN states.

\begin{figure}
    \centering
    \includegraphics[width=\linewidth]{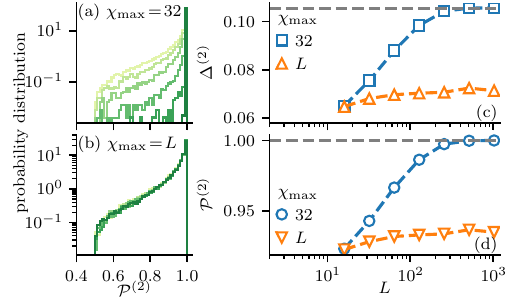}
    \caption{Results for the RTN states of the form in Eq.~\eqref{eq:rtn-post-meas}. (a)-(b) Distributions of the bipartite purity, ${\cal P}^{(2)}$ between $R_1$ and $R_2$ over several instances of $\ket{\tilde{\psi}_{\rm RTN}}$, for maximum bond dimension, $\chi_{\rm max}$, (a) finite and (b) equal to $L$. Data are for $L=32,64,128,256,512,1024$ where darker colours denote larger $L$. 
    (c) Average ${\cal P}^{(2)}$ and (d) Average $\Delta^{(2)}$ as a function of $L$. The grey dashed lines denote the corresponding values when $R_1$ and $R_2$ are decoupled and described by independent Gaussian random states.}
    \label{fig:rtn}
\end{figure}

We consider two specific cases. First, as a proxy for finite-depth circuits, we consider $\chi_i=\min[\chi_{\rm max},D_i]$ where $\chi_{\rm max} = e^{v_{\rm E}t}\sim O(1)$.
In this case, the distribution of ${\cal P}^{(2)}$ shifts towards a Dirac-$\delta$ function at 1 with increasing $L$ as shown in Fig.~\ref{fig:rtn}(a).
Concomitantly, ${\cal P}^{(2)}$ and $\Delta^{(2)}$ also grow with $L$ saturating to values corresponding to the case where in the RTN state $R_1$ and $R_2$ are decoupled from each other as evident in Fig.~\ref{fig:rtn}(c)-(d). 
This corroborates the result that there is no deep thermalisation or entanglement teleportation at finite times.

Second, to mimic circuits with depth $\propto \ln L$, we consider $\chi_i=\min[L,\ln D_i]$, in other words, $\chi_{\rm max}=L$. 
In this case the distribution of ${\cal P}^{(2)}$ is converged with $L$, as seen in Fig.~\ref{fig:rtn}(b). Consistently, ${\cal P}^{(2)}$ and $\Delta^{(2)}$, shown in Fig.~\ref{fig:rtn}(c)-(d), are also independent of $L$ and saturate to values smaller than the corresponding values for the decoupled case. This indicates finite entanglement between $R_1$ and $R_2$ as well as the onset of deep thermalisation. 
This shows that for finite entanglement to develop between $R_1$ and $R_2$, the circuit must be of depth $\sim \ln L$ which is consistent with the results discussed earlier for the brickwork circuit.  

To obtain analytical insights into this result, we make two simplifying approximations.
First, we consider the bond dimension to be uniform throughout the state in Eq.~\eqref{eq:rtn-post-meas}, such that each $\tilde{M}^{[i]}$ is a $\chi\times\chi$ Gaussian random tensor except $\tilde{M}^{[0]}$ and $\tilde{M}^{[L+1]}$, which are $2\times \chi$ tensors.
Second, we consider the annealed purity
\eq{
{\cal P}_{\rm ann} = \frac{\mathbb{E}[{\rm Tr}_{R_1}[{\rm Tr}_{R_2}\ket{\tilde{\psi}_{\rm RTN}}\bra{\tilde{\psi}_{\rm RTN}}]^2]}{\mathbb{E}\left[\braket{\tilde{\psi}_{\rm RTN}|\tilde{\psi}_{\rm RTN}}^2\right]}\,,
\label{eq:purity-ann}
}
where $\mathbb{E}[\cdot]$ denotes the average over the RTNs. 
Using standard rules for averaging Gaussian random tensors, we find that (see SM for details~\cite{supp})
\eq{
{\cal P}_{\rm ann} = \frac{9-Q(\chi,L)}{9+Q(\chi,L)}\,;~~Q(\chi,L) = \left(\frac{\chi-1}{\chi+1}\right)^{L+1}.
\label{eq:Pann-res}
}
The above can be used to show that for there to be any finite entanglement between $R_1$ and $R_2$ (${\cal P}_{\rm ann}<1$ strictly) for $L\gg 1$, $\chi$ must be $\propto L$. Any slower growth of $\chi$ with $L$ results in ${\cal P}_{\rm ann}=1$ in the $L\to\infty$ limit leading to vanishing entanglement. 
Since the effective bond dimension due to scrambling dynamics grows exponentially with the depth of the circuit, it naturally means that the depth must be $\propto \ln L$ for $\chi$ to be $\propto L$.
This concludes the argument that, in locally interacting circuits, there is a timescale proportional to the logarithm of the separation between $R_1$ and $R_2$ for entanglement teleportation and, by the bound in Eq.~\eqref{eq:delk-puk-ineq}, for deep thermalisation as well.

\paragraph*{Summary and Outlook:}
To summarise, we showed that deep thermalisation of a subsystem with disjoint, causally disconnected subregions is constrained by the measurement-induced entanglement teleportation between the subregions. 
For `generic' locally interacting quantum systems, this leads to timescale which grows logarithmically with the separation between the subregions for deep thermalisation, which is a manifestation of the emergent locality of deep thermalisation. 
In cases where the unitary circuit perfectly transmits the randomness of measurement outcomes to the ensemble of states on the subsystem, deep thermalisation can occur within $O(1)$ timescales, indicating non-locality of the measurement processes. The minimal model where the bath was modelled via Haar-random unitaries in infinitely large Hilbert-space dimensions is one such example. 
However, such situations may also arise in locally interacting models provided they are fine-tuned, such as in space-time self-dual models~\cite{ho2022exact,shrotriya2025non} or have infinite local Hilbert-space dimensions~\cite{chan2024pe}.

Our work also raises some natural questions. While we considered circuits with no conserved quantities, it will be interesting to check if circuits with conserved charges or effecting constrained dynamics impose further local structures on deep thermalisation and entanglement teleportation due to the diffusive behaviour of the conserved densities.
It will also be interesting to study the effect of partial measurements, such as where the measurement outcomes of a part of $E$ are discarded, on entanglement teleportation. More specifically, this may lead to mixed-state deep thermalisation (MSDT)~\cite{yu2025mixed,sherry2026deep} and it will be important to ask how mixed-state entanglement measures interplay with the locality, or lack thereof, in MSDT.

\paragraph*{Acknowledgements:} The authors thank W. W. Ho for several useful discussions. This work was supported by the Department of Atomic Energy, Government of India, under project nos. RTI4019 and RTI4013.
S.R. acknowledges support from SERB-DST (India) under Grant No. SRG/2023/000858, from ANRF (India) under Grant No. ANRF/ARG/2025/004045/PS, and from a Max Planck Partner Group grant between ICTS-TIFR, Bengaluru and MPIPKS, Dresden.

\bibliography{refs}

\clearpage

\setcounter{equation}{0}
\setcounter{figure}{0}
\setcounter{page}{1}
\renewcommand{\theequation}{S\arabic{equation}}
\renewcommand{\thefigure}{S\arabic{figure}}
\renewcommand{\thesection}{S\arabic{section}}
\renewcommand{\thepage}{S\arabic{page}}

\onecolumngrid
\begin{center}
{\bf {\large{Supplementary Material: \mytitle}}}\\
\medskip

Saptarshi Mandal, Alan Sherry, and Sthitadhi Roy\\
{\it {\small{International Centre for Theoretical Sciences, Tata Institute of Fundamental Research, Bengaluru 560089, India}}}
\end{center}
\twocolumngrid
\section{Minimal model with a dual-unitary gate at the bottleneck}

In this section, we show how having just one of the gates between $S_1$ and $E_1$, and between $S_2$ and $E_2$ to be dual-unitary in the minimal model in Eq.~\eqref{eq:toy-model-unitary} leads to perfect deep thermalisation. 
To show this we compute $\rho_{R}^{(k)}$ and show that it is identical to that of the Haar ensemble. 
The moment in Eq.~\eqref{eq:toy-model-moment} can be evaluated using the replica trick as~\cite{Claeys2022emergent}
\eq{
\rho_{R}^{(k)}
=
\lim_{n\to1-k}
{\rm Tr}^{\otimes n}_{R}
\Phi(n,k),
\label{eq:replica-to-real-moment}
}
where $\Phi(n,k)$ can be interpreted as a density matrix defined on $m=(n+k)$ replicas of the Hilbert space of $R$, and ${\rm Tr}^{\otimes n}$ denotes the trace over $n$ of the copies.
Using the standard toolbox of Weingarten calculus~\cite{weingarten1978asymptotic}, the average of $\phi$ over the Haar ensemble in Eq.~\eqref{eq:toy-model-moment} can be evaluated, which gives
\eq{
\Phi(n,k)=
\frac{\sqrt{D_\phi}(D_\phi-1)!}{(D_\phi+m-1)!}
\sum_{\substack{
\sigma\in \mathsf{S}_{m},\\
z_{S}
}}
{\cal D}_{\sigma}(z_{S})\,,
\label{eq:Phi-nk-def}
}
where $\mathsf{S}_m$ denotes the symmetric group and $D_\sigma(z_S)$ is given by
\eq{
    {\cal D}_{\sigma}(z_S) = \cbox{
    \begin{tikzpicture}[scale=0.75]
% outer legs
\foreach \x in {-2.7,-1.8,1.8,2.7}{
    \draw[thick] (\x,0)--(\x,5.8);

    \filldraw[
    fill=white,
    draw=black,
    thick]
    (\x-0.15,0)
    --
    (\x+0.15,0)
    --
    (\x,-0.2)
    --cycle;
}

% top projectors
\foreach \x in {-1.8,1.8}{
    \filldraw[
    fill=Green!40,
    draw=black,
    thick]
    (\x-0.15,3.2+2.4)
    --
    (\x+0.15,3.2+2.4)
    --
    (\x,3.4+2.4)
    --cycle;
}

\node at (-1.6,3.8+2.4)
{\small $\Pi_{z_{S_1}}^{\otimes(n+k)}$};

\node at (2.0,3.8+2.4)
{\small $\Pi_{z_{S_2}}^{\otimes(n+k)}$};

% sigma contractions
\foreach \i in {-0.2,1,2.2,3.4}{

\draw[
thick,
rounded corners]
(-1.0,1.5+\i)
--++(0,0.3)
--++(0.35,0);

\filldraw[
thick,
fill=white]
(-0.5,1.8+\i)
circle (0.18);

\node at (-0.5,1.8+\i)
{\small $\sigma$};

\draw[
thick,
rounded corners]
(-1.0,1.5+\i)
--++(0,-0.3)
--++(0.35,0);

\filldraw[
thick,
fill=white]
(-0.5,1.2+\i)
circle (0.18);

\node at (-0.5,1.2+\i)
{\small $\sigma$};

\draw[
thick,
rounded corners]
(1.0,1.5+\i)
--++(0,0.3)
--++(-0.35,0);

\filldraw[
thick,
fill=white]
(0.5,1.8+\i)
circle (0.18);

\node at (0.5,1.8+\i)
{\small $\sigma$};

\draw[
thick,
rounded corners]
(1.0,1.5+\i)
--++(0,-0.3)
--++(-0.35,0);

\filldraw[
thick,
fill=white]
(0.5,1.2+\i)
circle (0.18);

\node at (0.5,1.2+\i)
{\small$\sigma$};
}

% purple blocks
\foreach \y in {0.8,2,3.2,4.4}{

\filldraw[
fill=NavyBlue!80,
draw=black,
rounded corners,
thick]
(-2.0,\y+0.3)
rectangle
(-0.8,\y+0.7);
% \node at (-1.4,\y+0.5) {\small $\otimes(n+k)$}; 
\filldraw[
fill=NavyBlue!80,
draw=black,
rounded corners,
thick]
(0.8,\y+0.3)
rectangle
(2.0,\y+0.7);
}
\filldraw[
fill=OliveGreen!80,
draw=black,
rounded corners,
thick]
(0.8,2+0.3)
rectangle
(2.0,2+0.7);

\filldraw[
fill=OliveGreen!80,
draw=black,
rounded corners,
thick]
(-2.0,3.2+0.3)
rectangle
(-0.8,3.2+0.7);

% gray blocks
\filldraw[
fill=gray!50,
draw=black,
rounded corners,
thick]
(-2.95,0.3)
rectangle
(-1.55,0.7);

\filldraw[
fill=gray!50,
draw=black,
rounded corners,
thick]
(1.55,0.3)
rectangle
(2.95,0.7);

\end{tikzpicture}
    }\,,
    \label{eq:toy-model-Dsigma}
}
where each box now represents $m=n+k$ replicas of the gate and the green gates are dual unitary.
Dual unitarity simply means that a gate continues to be unitary under exchange of space and time~\cite{bertini2025exactlysolvablemanybodydynamics}.
Diagrammatically this implies 
\eq{
\cbox{
\begin{tikzpicture}[scale=0.75]
\foreach \i in {-0.2}{
\draw[thick,rounded corners](-1.0,1.5+\i)--++(0,0.3)--++(0.35,0);
\filldraw[thick,fill=white](-0.5,1.8+\i)circle (0.18);
\node at (-0.5,1.8+\i){\small $\sigma$};
\draw[thick,rounded corners](-1.0,1.5+\i)--++(0,-0.3)--++(0.35,0);
\filldraw[thick,fill=white](-0.5,1.2+\i)circle (0.18);
\node at (-0.5,1.2+\i){\small $\sigma$};
}
\foreach \y in {0.8}{
\draw[thick] (-1.8,\y+0.3-0.2)--(-1.8,\y+0.7+0.2);
\filldraw[fill=OliveGreen!80,draw=black,rounded corners,thick](-2.0,\y+0.3) rectangle (-0.8,\y+0.7);
}
\end{tikzpicture}
}
=
\cbox{
\begin{tikzpicture}[scale=0.75]
\foreach \i in {-0.2}{
\draw[thick,rounded corners](-0.5,1.8+\i) -- ++(0,0.4);
\filldraw[thick,fill=white](-0.5,1.8+\i)circle (0.18);
\node at (-0.5,1.8+\i){\small $\sigma$};
\draw[thick,rounded corners](-0.5,1.2+\i) -- ++(0,-0.4);
\filldraw[thick,fill=white](-0.5,1.2+\i)circle (0.18);
\node at (-0.5,1.2+\i){\small $\sigma$};
}
\end{tikzpicture}
}\,.
\label{eq:du-contract}
}
In addition, we will use the fact that the various contractions give 
\eq{
\cbox{
\begin{tikzpicture}[scale=0.75]
\foreach \i in {0}{
    \draw[thick,rounded corners] (-2,1.5 + \i) -- ++(0,0.3) -- ++(0.6,0);
    \filldraw[thick,fill=white] (-2+0.6+0.15,1.5 +\i +0.3) circle (0.18);
    \node at (-2+0.6+0.15,1.5 +\i +0.3) {\small $\sigma$};
    \draw[thick,rounded corners] (-2,1.5 +\i) -- ++(0,-0.3) -- ++(0.6,0);
    \filldraw[thick,fill=white] (-2+0.6+0.15,1.5 +\i -0.3) circle (0.18);
    \node at (-2+0.6+0.15,1.5 +\i -0.3) {\small $\sigma$};
    }
\end{tikzpicture}
} = D_{E_{1/2}}^{n+k}~~~~{\rm and}~~~~
\cbox{
\begin{tikzpicture}[scale=0.75]
\foreach \i in {0}{
    \draw[thick,rounded corners] (-2,1.5 + \i) -- ++(0,0.3) -- ++(0.6,0);
    \filldraw[thick,fill=white] (-2+0.6+0.15,1.5 +\i +0.3) circle (0.18);
    \node at (-2+0.6+0.15,1.5 +\i +0.3) {$\sigma$};
    \draw[thick,rounded corners] (-2,1.5 +\i) -- ++(0,-0.3) -- ++(0.6,0);
    \filldraw[thick,fill=Green!40] (-1.4,1.2-0.15) -- (-1.4,1.2+0.15) -- (-1.4+0.15,1.2) -- cycle;
    }
\end{tikzpicture}
}=1\,,
\label{eq:perm-contract}
}
where the second equality follows from the idempotency of the projection operator.
Also, note that the grey gates between $S_{1/2}$ and $E_{1/2}$ denote $n+k$ copies of a Bell state between them which leads to the useful identity
\eq{
\cbox{
\begin{tikzpicture}[scale=0.75]
\foreach \x in {1,2}{
    \draw[thick](\x,0) -- (\x,1);
    \filldraw[fill=white, draw=black, thick] (\x-0.15,0) -- (\x+0.15,0) -- (\x,-0.2) -- cycle;
}
    \filldraw [fill=Gray!50, draw=black,rounded corners, thick] (0.8,0.3) rectangle (2.2,0.7);
    ;
\filldraw[thick, fill=white] (2,1.15) circle (0.18);
    \node at (2,1.15) {\small $\sigma$};
\end{tikzpicture}
}
=
2^{-(n+k)}
\cbox{
\begin{tikzpicture}[scale=0.75]
\draw[thick] (0,-1.) -- (0,-0.0);
\filldraw[thick, fill=white] (0,-1) circle (0.18);
\node at (0,-1) {\small $\sigma$};
\end{tikzpicture}
}
\label{eq:bell-perm-contraction}
}
in the $(n+k)$-copy folded picture.

Using Eqs.~\eqref{eq:du-contract}, \eqref{eq:perm-contract}, and \eqref{eq:bell-perm-contraction} in Eq.~\eqref{eq:toy-model-Dsigma}, together with $D_{S_{1/2}}=D_{E_{1/2}}=2$, one obtains for arbitrary $t$
\eq{
{\cal D}_\sigma(z_S)= 2^{2m(t-2)}
\cbox{
\begin{tikzpicture}[scale=0.75]
\draw[thick] (0,-1) -- (0,-0.5);
\draw[thick] (0.5,-1) -- (0.5,-0.5);
\filldraw[thick, fill=white] (0,-1.15) circle (0.18);
\filldraw[thick, fill=white] (0.5,-1.15) circle (0.18);
\node at (0,-1.15) {\small $\sigma$};
\node at (0.5,-1.15) {\small $\sigma$};
\end{tikzpicture}
}\,,
\label{eq:phink-simplified}
}
from which $\Phi(n,k)$ can be straightforwardly obtained using Eq.~\eqref{eq:Phi-nk-def}.

The remaining task is to trace out the $n$ replica copies as prescribed by Eq.~\eqref{eq:replica-to-real-moment}. To this end, we use the identity
\eq{
\sum_{\sigma\in \mathsf{S}_{m}}{\rm Tr}^{\otimes n}\pi_{\sigma}=\frac{(d+m-1)!}{(d-1)!}\rho^{(k)}_{\rm Haar}(d)\,.
\label{eq:replica-trace-identity}
}
We therefore have
\eq{
{\rm Tr}_R^{\otimes n}\Phi(n,k) = \rho_{\rm Haar}^{(k)}&(D_R)\times 2^{2m(t-2)+2}\times\nonumber\\
&\frac{(D_R+m-1)!}{(D_R-1)!}
\frac{\sqrt{D_\phi}(D_\phi-1)!}{(D_\phi+m-1)!}\,
}
where we can take the replica limit, $n\to 1-k$, such that
\eq{
\rho^{(k)}_{R} = \rho_{\rm Haar}^{(k)}(D_R)2^{2(t-2)+2}\times D_R\times D_\phi^{-1/2}\,.
}
Using $D_\phi=D_{E_1}^{2t}D_{E_2}^{2t}=2^{4t}$ and $D_R=4$ in the above, we obtain
\eq{
\rho^{(k)}_{R}=\rho^{(k)}_{\rm Haar}(D_R)\,.
}
This proves that ${\cal E}_R$ is given exactly by the Haar ensemble. 
The independence of these results on $t$ demonstrates that the model exhibits complete teleportation of entanglement and perfect deep thermalisation for any $t$ provided there exists at least one dual-unitary gate between $S_1$ and $E_1$, and between $S_2$ and $E_2$.

\section{Additional numerical results for brickwork circuit}
In this section, we present additional numerical results for the brickwork circuit introduced in the main text, focusing on the $k$-dependence of $\delta {\cal P}^{(k)}$ and $\Delta^{(k)}$. In the main text, Fig.~\ref{fig:ising-plot} establishes the teleportation timescale using $k=4$ across different system sizes. Here we show results for several values of $k$ in order to examine whether there exists a hierarchy of timescales with $k$.
\begin{figure}
\centering
\includegraphics[width=\linewidth]{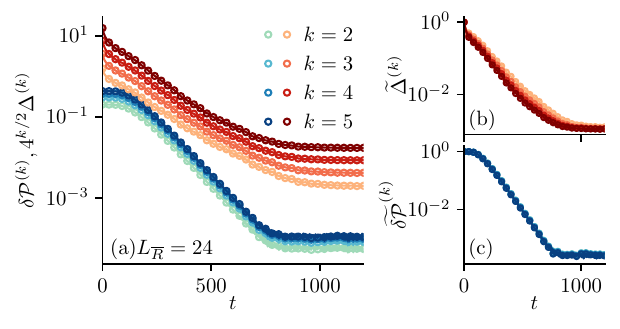}
\caption{(a) Numerical results for $\delta {\cal P}^{(k)}$ and $\Delta^{(k)}$ for the brickwork circuit in Eqs.~\eqref{eq:brickwork}--\eqref{eq:ham}, shown here for $L_{\overline R}=24$. Darker colours correspond to larger $k$. The data are consistent with the bound in Eq.~\eqref{eq:delk-puk-ineq}. (b) The normalized quantity $\widetilde{\Delta}^{(k)}(t)=\Delta^{(k)}(t)/\Delta^{(k)}(0)$ exhibits a $k$-dependence in its decay. (c) By contrast, $\widetilde{\delta {\cal P}}^{(k)}(t)=\delta {\cal P}^{(k)}(t)/\delta {\cal P}^{(k)}(0)$ displays a universal exponential decay as a function of $t/t_\ast(\lrb)$ for $t\gtrsim t_\ast$ with a $k$-independent rate.}
\label{fig:ising-k-plot}
\end{figure}

The numerical data for $L_{\overline R}=24$ are shown in Fig.~\ref{fig:ising-k-plot}. As in the main text, the data are consistent with the bound in Eq.~\eqref{eq:delk-puk-ineq}. However, $\delta {\cal P}^{(k)}$ and $\Delta^{(k)}$ exhibit qualitatively different $k$-dependence. In particular, we do not observe any hierarchy of timescales in $\delta {\cal P}^{(k)}$. Instead, the normalized quantity
\eq{
\widetilde{\delta {\cal P}}^{(k)}(t)
\equiv
\frac{\delta {\cal P}^{(k)}(t)}{\delta {\cal P}^{(k)}(0)}
}
collapses onto a single curve for different $k$ when plotted as a function of $t/t_\ast$, where $t_\ast$ is the teleportation timescale extracted in the main text. The data are consistent with the scaling form
\eq{
\widetilde{\delta {\cal P}}^{(k)}(t)=f(t/t_\ast),
}
where $f(x)\approx 1$ for $x\ll 1$, decays approximately as $e^{-\lambda x}$ for $x\gtrsim 1$, and saturates to a value which itself is exponentially small in $\lrb$ at late times.
The key point is that the $\lambda$, being independent of $k$, suggests that entanglement teleportation, as quantified by the $k^{\rm th}$-R\'enyi entropy of entanglement, happens over the same timescale for all $k$.

By contrast, the normalized quantity
\eq{
\widetilde{\Delta}^{(k)}(t)
\equiv
\frac{\Delta^{(k)}(t)}{\Delta^{(k)}(0)}
}
shows a visible dependence on $k$, as illustrated in Fig.~\ref{fig:ising-k-plot}(b). Thus, while the teleported entanglement, quantified by $\delta {\cal P}^{(k)}$ is governed by a universal decay profile independent of $k$, the decay of $\Delta^{(k)}$ retains additional $k$-dependent structure. Together with the inequality in Eq.~\eqref{eq:delk-puk-ineq}, this suggests that the dynamics of $\Delta^{(k)}$ is not determined solely by the teleported entanglement captured by $\delta {\cal P}^{(k)}$, but also receives contributions from other correlations. These additional contributions can then generate a hierarchy of deep-thermalisation timescales, or equivalently of the design times discussed in~\cite{ippoliti2022solvablemodelofdeep,cotler2023emergent}.

\section{Details of bipartite purity of RTN state}
In this section, we present the details of the analytic computation of ${\cal P}_{\rm ann}$ for RTN states constructed from Gaussian random tensors of uniform bond dimension $\chi$. In order to do this, it will be useful to interpret the state in Eq.~\eqref{eq:rtn-post-meas} as a $2\times 2$ tensor
\eq{
\Phi = \cbox{
\begin{tikzpicture}[scale=1]
\draw[thick] (0,0) -- (2,0);
\draw[thick] (3.5,0) -- (4.5,0);
\draw[thick,dashed] (2,0) -- (4,0);
\draw[black,thick] (0,0) -- (0,0.5);
\draw[black,thick] (4.5,0) -- (4.5,0.5);
\foreach \x in {0,1,2,3.5,4.5}{
    \filldraw[draw=black,thick,fill=Blue!10, rounded corners] (\x-0.25,-0.25) rectangle (\x+0.25,0.25);
}
\node at (0,-0.5) {\small $\tilde{M}^{[0]}$};
\node at (1,-0.5) {\small $\tilde{M}^{[1]}$};
\node at (2,-0.5) {\small $\tilde{M}^{[2]}$};
\node at (3.5,-0.5) {\small $\tilde{M}^{[L]}$};
\node at (4.5,-0.5) {\small $\tilde{M}^{[L+1]}$};
\end{tikzpicture}
}\,.
}
With this, Eq.~\eqref{eq:purity-ann} can be recast as 
\eq{
{\cal P}_{\rm ann} = \frac{{\cal N}}{{\cal D}} = \frac{ \mathbb{E}[{\rm Tr}((\Phi \Phi^\dagger)^2)]}{\mathbb{E} [{\rm Tr}(\Phi^\dagger \Phi)]^2}\,,
\label{eq:Pann-frac}
}
Using the diagrammatic notation
\eq{
{\cal T} \equiv 
\cbox{
\begin{tikzpicture}[scale=1]
\draw[very thick] (-0.5,0) -- (0.5,0);
\filldraw[draw=black,very thick,fill=BrickRed!50, rounded corners] (-0.25,-0.25) rectangle (0.25,0.25);  
\end{tikzpicture}
}=
\cbox{
\begin{tikzpicture}[scale=1]
\draw[thick] (-0.5,0) -- (0.5,0);
\filldraw[draw=black,thick,fill=Red!10, rounded corners] (-0.25,-0.25) rectangle (0.25,0.25);    
\draw[thick] (-0.5-0.1,0-0.2) -- (0.5-0.1,0-0.2);
\filldraw[draw=black,thick,fill=Blue!10, rounded corners] (-0.25-0.1,-0.25-0.2) rectangle (0.25-0.1,0.25-0.2);    
\draw[thick] (-0.5-0.2,0-0.4) -- (0.5-0.2,0-0.4);
\filldraw[draw=black,thick,fill=Red!10, rounded corners] (-0.25-.2,-0.25-.4) rectangle (0.25-0.2,0.25-0.4);    
\draw[thick] (-0.5-0.3,0-0.6) -- (0.5-0.3,0-0.6);
\filldraw[draw=black,thick,fill=Blue!10, rounded corners] (-0.25-0.3,-0.25-0.6) rectangle (0.25-0.3,0.25-0.6);    
% \draw[thick] (-0.5,0) -- (0.5,0);
% \filldraw[draw=black,thick,fill=Red!10, rounded corners] (-0.25,-0.25) rectangle (0.25,0.25);    
% \draw[thick] (-0.5,0) -- (0.5,0);
% \filldraw[draw=black,thick,fill=Red!10, rounded corners] (-0.25,-0.25) rectangle (0.25,0.25);    
\end{tikzpicture}
}\,,
\label{eq:def-T}
}
where the light blue and light red squares denote the tensors $M$ and $M^\ast$, respectively, 
the numerator and denominator in Eq.~\eqref{eq:Pann-frac} can be expressed as 
\eq{
\begin{split}
{\cal N} = \mathbb{E}\left[\cbox{
\begin{tikzpicture}[scale=1]
\draw[very thick] (0,0) -- (2,0);
\draw[very thick] (3.5,0) -- (4.5,0);
\draw[very thick,dashed] (2,0) -- (4,0);
\draw[black,very thick] (0,0) -- (0,0.5);
\draw[black,very thick] (4.5,0) -- (4.5,0.5);
\filldraw[draw=black, very thick, fill=white] (-0.1,0.5) rectangle (0.1,0.7);
\filldraw[draw=black, very thick, fill=white] (4.5,0.6) circle (0.1);
\foreach \x in {0,1,2,3.5,4.5}{
    \filldraw[draw=black,very thick,fill=BrickRed!50, rounded corners] (\x-0.25,-0.25) rectangle (\x+0.25,0.25);
}
\end{tikzpicture}
}\right]\,,\\
{\cal D} = \mathbb{E}\left[\cbox{
\begin{tikzpicture}[scale=1]
\draw[very thick] (0,0) -- (2,0);
\draw[very thick] (3.5,0) -- (4.5,0);
\draw[very thick,dashed] (2,0) -- (4,0);
\draw[black,very thick] (0,0) -- (0,0.5);
\draw[black,very thick] (4.5,0) -- (4.5,0.5);
\filldraw[draw=black, very thick, fill=white] (0,0.6) circle (0.1);
\filldraw[draw=black, very thick, fill=white] (4.5,0.6) circle (0.1);
\foreach \x in {0,1,2,3.5,4.5}{
    \filldraw[draw=black,very thick,fill=BrickRed!50, rounded corners] (\x-0.25,-0.25) rectangle (\x+0.25,0.25);
}
\end{tikzpicture}
}\right]\,,
\end{split}
}
where 
$\cbox{\tikz{
\draw[very thick] (4.2,0.6) -- (4.4,0.6);
\filldraw[draw=black, very thick, fill=white] (4.5,0.6) circle (0.1);}}\equiv|\mathbb{I})$
denotes the identity permutation and 
$\cbox{\tikz{\draw[black,very thick] (-0.3,0.6) -- (-0.1,0.6);
\filldraw[draw=black, very thick, fill=white] (-0.1,0.5) rectangle (0.1,0.7);}}\equiv|\mathbb{S})$
denotes the swap permutation between the two replicas of $M$ and $M^\ast$ in the definition of ${\cal T}$ in Eq.~\eqref{eq:def-T}.
Using the standard rules of averaging over Gaussian random tensors, we have
\eq{
\overline{{\cal T}}\equiv \mathbb{E}[{\cal T}] = \cbox{\tikz{
\draw[very thick] (4.2,0.6) -- (4.4,0.6);
\filldraw[draw=black, very thick, fill=white] (4.5,0.6) circle (0.1);}}~\cbox{\tikz{
\draw[very thick] (4.6,0.6) -- (4.8,0.6);
\filldraw[draw=black, very thick, fill=white] (4.5,0.6) circle (0.1);}}
+
\cbox{\tikz{\draw[black,very thick] (-0.3,0.6) -- (-0.1,0.6);
\filldraw[draw=black, very thick, fill=white] (-0.1,0.5) rectangle (0.1,0.7);}}~
\cbox{\tikz{\draw[black,very thick] (0.1,0.6) -- (0.3,0.6);
\filldraw[draw=black, very thick, fill=white] (-0.1,0.5) rectangle (0.1,0.7);}}\,.
}
In addition, contractions between these permutations give
\eq{
\cbox{\tikz{
\draw[very thick] (4.6,0.6) -- (4.8,0.6);
\filldraw[draw=black, very thick, fill=white] (4.5,0.6) circle (0.1);}}\!
\cbox{\tikz{
\draw[very thick] (4.2,0.6) -- (4.4,0.6);
\filldraw[draw=black, very thick, fill=white] (4.5,0.6) circle (0.1);}}
=
\cbox{\tikz{\draw[black,very thick] (0.1,0.6) -- (0.3,0.6);
\filldraw[draw=black, very thick, fill=white] (-0.1,0.5) rectangle (0.1,0.7);}}\!
\cbox{\tikz{\draw[black,very thick] (-0.3,0.6) -- (-0.1,0.6);
\filldraw[draw=black, very thick, fill=white] (-0.1,0.5) rectangle (0.1,0.7);}}=\chi^2;~~~
\cbox{\tikz{
\draw[very thick] (4.6,0.6) -- (4.8,0.6);
\filldraw[draw=black, very thick, fill=white] (4.5,0.6) circle (0.1);}}
\!
\cbox{\tikz{\draw[black,very thick] (-0.3,0.6) -- (-0.1,0.6);
\filldraw[draw=black, very thick, fill=white] (-0.1,0.5) rectangle (0.1,0.7);}}=\chi\,.
}
We can therefore express the transfer matrix between the permutations as 
\eq{
\overline{\cal T}=\begin{pmatrix}\chi^2&\chi\\\chi&\chi^2\end{pmatrix}\,,
\label{eq:T-mat}
}
such that 
\eq{
{\cal P}_{\rm ann} = \frac{\big(\mathbb{S}\big|{\overline{\cal T}}^{L+1}\big|\mathbb{I}\big)}{\big(\mathbb{I}\big|{\overline{\cal T}}^{L+1}\big|\mathbb{I}\big)}\,.
}
Note that at the boundaries, the bond dimensions are $\chi_b=2$, and therefore, in the same basis as in Eq.~\eqref{eq:T-mat}, the boundary contractions can be written as 
\eq{
|\mathbb{I}) = \begin{pmatrix}\chi_b^2\\\chi_b\end{pmatrix}=\begin{pmatrix}4\\2\end{pmatrix}\,;~
|\mathbb{S}) = \begin{pmatrix}\chi_b\\\chi_b^2\end{pmatrix}=\begin{pmatrix}2\\4\end{pmatrix}\,.
\label{eq:pann-tmat}
}
The transfer matrix in Eq.~\eqref{eq:T-mat} has eigenvalues and eigenvectors given by
\eq{
\lambda_\pm = \chi(\chi\pm 1)\;;~~\ket{\pm} = \frac{1}{\sqrt{2}}\begin{pmatrix}1\\\pm1\end{pmatrix}\,,
\label{eq:Tmat-eig}
}
and the boundary vectors in terms of these are given by
\eq{
\begin{split}
    |\mathbb{I}) = 3\sqrt{2} \ket{+} + \sqrt{2} \ket{-}\\
    |\mathbb{S}) = 3\sqrt{2} \ket{+} - \sqrt{2} \ket{-}
\end{split}\,.
\label{eq:bdry-eig}
}
Using Eqs.~\eqref{eq:Tmat-eig} and \eqref{eq:bdry-eig} in Eq.~\eqref{eq:pann-tmat}, we obtain
\eq{
{\cal P}_{\rm ann} = \frac{9 \lambda_+^{L+1} -  \lambda_-^{L+1}}{9 \lambda_+^{L+1} +  \lambda_-^{L+1}} = \frac{9-Q(\chi,L)}{9+Q(\chi,L)}\,,
}
with 
\eq{
Q(\chi,L) = \left(\frac{\chi-1}{\chi+1}\right)^{L+1}\,.
}
This is exactly the result in Eq.~\eqref{eq:Pann-res}.

Note that for any $O(1)$ value of $\chi$, $Q(\chi,L\to\infty)\to 0$ such that ${\cal P}_{\rm ann}\to 1$.
This indicates that at any finite depth, there is no entanglement between $R_1$ and $R_2$ in the thermodynamic limit. For $L\gg 1$, $Q(\chi,L)\ll 1$ such that we have 
\eq{
{\cal P}_{\rm ann} \approx 1-\frac{2}{9}\left(\frac{\chi-1}{\chi+1}\right)e^{-L\mathcal{X}}\,,
}
where $\mathcal{X} = \ln[(\chi+1)/(\chi-1)]$. This is consistent with the numerical result discussed in the main text.

By contrast, consider the case where $\chi=\alpha L$, which corresponds to a depth of the circuit $\propto \ln L$. 
In this case, 
\eq{
\lim_{L\to\infty}Q(\chi=\alpha L,L) = e^{-2/\alpha},
}
such that 
\eq{
{\cal P}_{\rm ann}= \frac{1-e^{-2/\alpha}/9}{1+e^{-2/\alpha}/9}\,,
}
which continues to be finite in the thermodynamic limit. 
This again shows that a depth $\propto \ln L$ is necessary for entanglement teleportation between $R_1$ and $R_2$.

\end{document}